\documentclass[aip,reprint,jap,groupedaddress]{revtex4-1}

\usepackage{graphics,color,url,hyperref}

\begin{document}

\title{Optimal uni-axial ferromagnetism in (La,Ce)$_2$Fe$_{14}$B for permanent magnets}

\author{
Munehisa~Matsumoto$^{1}$, Masaaki~Ito$^{2}$,
Noritsugu~Sakuma$^2$, Masao~Yano$^2$, Tetsuya~Shoji$^2$,
Hisazumi~Akai$^{1,3}$
}

\affiliation{
$^1$Institute for Solid State Physics, University of Tokyo, Kashiwa 277-8581, JAPAN\\
$^2$Advanced Material Engineering Div., Toyota~Motor~Corporation, Susono 410-1193, JAPAN\\
$^3$Elements Strategy Initiative Center for Magnetic Materials (ESICMM),
National Institute for Materials Science (NIMS), Sengen 1-2-1, Tsukuba 305-0047, JAPAN
}

\date{\today}

\begin{abstract}
Prospects for light-rare-earth-based permanent magnet compound R$_{2}$Fe$_{14}$B
(R=La$_{1-x}$Ce$_{x}$ with $0 \le x\le 1$)
are inspected from first principles referring to the latest experimental data.
Ce-rich 2:14:1 compounds come with good structure stability, reasonably good combination
of magnetization and magnetic anisotropy, while a drawback lies in the low Curie temperature
that is only 120~K above the room temperature at the Ce$_2$Fe$_{14}$B limit.
Best compromise is inspected on the basis of {\it ab initio} data for (La$_{1-x}$Ce$_{x}$)$_2$Fe$_{14}$B
referring to the magnetic properties of the champion
magnet compound Nd$_{2}$Fe$_{14}$B and prerequisite conditions imposed by practical utility.
\end{abstract}


%
%

\maketitle

\section{Introduction}
Rare-earth permanent magnets make one of the most important materials
in the upcoming decades with expectedly extensive industrial applications
in electric vehicles and robotics. Permanent magnets exploit a metastable
state in the magnetization curve of uni-axial ferromagnets.
Strong magnetization and high Curie temperature can be basically achieved
with Fe-based ferromagnets which can be supplemented with rare metals like Nd and Co
to get a good combination of uni-axial magnetic anisotropy and even higher Curie temperature,
respectively. Today's commercial champion magnet is made of Nd$_{2}$Fe$_{14}$B~\cite{sagawa_1984, croat_1984, rmp_1991}.
The problem with this material sometimes happens with relatively low Curie temperature
at 585~K. Addition of Co or Dy has been a practical solution to supplement high-temperature performance
and recent developments have been concentrated on finding a way to reduce the amount of such expensive
elements. Even Nd might face a short supply in the upcoming few decades and we are motivated to find a way
to reduce the amount of Nd. Thus the target compound in the present study is
(La$_{1-x}$Ce$_{x}$)$_2$Fe$_{14}$B
with $0\le x\le 1$ with the cheap light rare earth elements, La and Ce, as a possible replacement for
Nd$_{2}$Fe$_{14}$B.

We address the intrinsic magnetic properties of (La$_{1-x}$Ce$_{x}$)$_2$Fe$_{14}$B from first principles
as a complementary approach to the recent experimental study~\cite{apl_2016}
and estimate the merit of the compound as a function of $x$ referring to the intrinsic properties
of Nd$_2$Fe$_{14}$B. Addition of Co is excluded here
in order to elucidate the $4f$-electron physics in La$_{1-x}$Ce$_{x}$
keeping the rest simple by restricting the scope of
the $3d$-electron part only
to Fe. Conventional prospect for Ce in the permanent magnet community
seems to have been bad because Ce is typically
detrimental to Curie temperature and hardly contributes to
magnetization. We identify other utility of Ce
in the structure stabilization and possibly also in the volume effect.
We will also see that
magnetic anisotropy coming from Ce can be superior to that from
$3d$-electrons, if not on a par with Nd anisotropy. Furthermore, exceptionally high Curie temperature
in Ce-based ferromagnet has been known in CeRh$_3$B$_2$ whose Curie temperature at 125~K is higher than
its Gd counterpart at 90~K~\cite{cerh3b2}. These point to the possibility for
more exploitation of subtlety in the interplay between
$f$-electrons and $d$-electrons in Ce-based intermetallics for permanent magnets.

The rest of this paper is organized as follows. In the next section we describe our {\it ab initio} methods.
In Sec.~\ref{sec::results} calculated results for formation energy,
magnetization, magnetic anisotropy,
and Curie temperature
are presented
and the merit associated with them are estimated. In Sec.~\ref{sec::discussions} some subtle points
in the utility of Ce are discussed. The last section is devoted for conclusions and outlook.

\section{Methods}
\label{sec::methods}

Intrinsic magnetic properties of (La$_{1-x}$Ce$_{x}$)$_2$Fe$_{14}$B
with
$0\le x\le 1$ are calculated from first principles combining two approaches.
One is {\it ab initio} structure optimization utilizing the open-source
software package OpenMX~\cite{OpenMX,Ozaki2003,Ozaki2004,Ozaki2005,Duy2014,Lejaeghere2016}
on the basis of pseudopotentials~\cite{MBK1993,Theurich2001}
and local orbital basis sets. This aproach is attempted
at the stoichiometric limits, $x=0$ and $x=1$,
and also for chemical compositions where discrete replacement is doable in the unit cell,
i.e. $x=n/8$ with $n=1,2,\ldots,7$,
out of the eight rare-earth atomic sites in the unit cell of 2:14:1 crystal structure~\cite{rmp_1991}.
The other is Korringa-Kohn-Rostoker (KKR)~\cite{korringa,kohn} Green's function approach
incorporating coherent potential approximation (CPA)~\cite{shiba_akai}
as implemented in the package AkaiKKR~\cite{AkaiKKR}.
With KKR-CPA, replacements between La and Ce can be explored continuously on the chemical
composition axis. The experimentally inspected mixing ratio and the particular
sublattice preference of Ce in La~\cite{apl_2016} can be continuously simulated as well.

\subsection{Discrete and continous interpolation {\it ab initio} between La$_2$Fe$_{14}$B and Ce$_{2}$Fe$_{14}$B}
\label{sec::methods::codes}

\subsubsection{{\it Ab initio} structure optimization: the discrete way}
\label{sec::methods::OpenMX}

In {\it ab initio} electronic structure calculations based on density functional theory~\cite{hohenberg_1964,kohn_1965},
it has been known that the lattice structure of Fe-based ferromagnets seems to be
best described within Generalized Gradient Approximation (GGA) as proposed
Perdew, Burke, and Ernzerhof (PBE)~\cite{prl_1996}. We do {\it ab initio} structure optimization
for (La,Ce)$_2$Fe$_{14}$B to track the trend of formation energy from La$_2$Fe$_{14}$B
to Ce$_2$Fe$_{14}$B. The basis set we take in OpenMX is \verb|Ce8.0-s2p2d2f1|,
\verb|La8.0-s2p2d2f1|,
\verb|Fe6.0S-s2p2d1|, and \verb|B7.0-s3p3d2| within the given pseudopotential data set~\cite{OpenMX}.
In the discussions below Y$_2$Fe$_{14}$B is sometimes taken as a $3d$-electron analogue
for the reference to elucidate the contribution from $4f$-electron part in La$_{1-x}$Ce$_x$.
For that case the basis \verb|Y8.0-s3p2d2f1| is used for Y$_2$Fe$_{14}$B. The energy cutoff is set to be 500~Ry
which was tuned together with the choice of the basis sets to ensure the convergence.
Too rich basis set can trigger instability the calculated results presumably due to overcompleteness of the basis~\cite{Ozaki2004}
and we did not quite achieve the convergence with richer basis sets like \verb|Ce8.0-s3p3d3f2| 
in the given pseudopotential data set~\cite{OpenMX} for this particular case
of Ce$_2$Fe$_{14}$B. For the present purpose
to systematically track the trends between La$_2$Fe$_{14}$B to Ce$_2$Fe$_{14}$B,
we settled down with the basis set \verb|Ce8.0-s2p2d2f1|. Similar basis sets with a little more or
less inclusion of local basis wavefunctions
can be good as well depending on the target materials
and the issue being investigated as long as the choice of the basis set
is coherently applied to the target materials and the reference materials.

Number of k points is set to be $64$ for the 2:14:1 materials.
For Ce$_2$Fe$_{14}$B, more extensive runs with k-point number being up to $256$
were verified as well and we have seen that sufficient convergence has been achieved
already at the stage where the k point number is $64$.

The starting structure
is taken from the experimental measurements~\cite{rmp_1991,apl_2016}
and {\it ab initio} structure optimization is done for (La$_{1-x}$Ce$_{x}$)$_2$Fe$_{14}$B with $x=n/8$ where
$n=0,1,2,\ldots,8$ to get the minimized energy $U_{\rm tot}[\mbox{(La$_{1-x}$Ce$_{x}$)$_2$Fe$_{14}$B}]$
and the associated magnetization in the ground state.
The structure optimization is done allowing for magnetic polarization without spin-orbit interaction.
In this way an optimized lattice and the associated magnetization is determined as a function of $x=n/8$
for (La$_{1-x}$Ce$_{x}$)$_2$Fe$_{14}$B discretely on the chemical composition axis.

\subsubsection{KKR-CPA on an interpolated lattice: the continuous way}
\label{sec::methods::AkaiKKR}

Continuous interpolation over $0\le x \le 1$ for (La$_{1-x}$Ce$_x$)$_2$Fe$_{14}$B
is done on the basis of KKR-CPA. 
Scalar relativistic calculations are done to be as realistic as possible to compare with
the experiments. We set $l_{\rm max}=3$ putting all $4f$ electrons in the valence state. This should be the proper description
for La and Ce$^{4+}$ as found in the $4f$-$3d$ intermetallics containing ferromagnetic Fe.
Possibility for Ce$^{3+}$ is discussed in Sec.~\ref{sec::discussions::valence} below.
This setup is in contrast to what is typically done for Nd where very well localized $4f$-electrons
are completely put outside of the valence state within the open-core approximation.

The lattice structure have to be fixed in the input
and we refer to the experimental lattice~\cite{rmp_1991,apl_2016} for the input.
Most typically in the present work we take the lattice constants of La$_2$Fe$_{14}$B at $x=0$
and those of Ce$_2$Fe$_{14}$B at $x=1$ from Ref.~\onlinecite{rmp_1991}, recycling the internal coordinates
of Nd$_2$Fe$_{14}$B for all $x$ in $0\le x\le 1$, and try to simulate the trends of intrinsic properties
in (La$_{1-x}$Ce$_x$)$_2$Fe$_{14}$B via ferromagnetic calculations with linearly interpolated lattice constants
\begin{eqnarray*}
a(x) & = & a[\mbox{La$_2$Fe$_{14}$B}]+x(a[\mbox{Ce$_2$Fe$_{14}$B}]-a[\mbox{La$_2$Fe$_{14}$B}])\\
c(x) & = & c[\mbox{La$_2$Fe$_{14}$B}]+x(c[\mbox{Ce$_2$Fe$_{14}$B}]-c[\mbox{La$_2$Fe$_{14}$B}])
\end{eqnarray*}
following Vegard's law. The strongest point of KKR-CPA lies in the flexibility where the
concentration of dopants and site segregation ratio can be manipulated continuously.
We thus fill in the data points on the $x$ axis that are not covered by OpenMX,
and we can look at the outcome of various ways of site segregation
including the particular way found experimentally~\cite{apl_2016}.

The exchange correlation is based on local spin density approximation
as parametrized by Moruzzi, Janak and Williams (MJW)~\cite{MJW} which is supposedly
one of the most established exchange correlations to describe the magnetism on a given lattice.

Care must be taken in the choice of muffin-tin radius or atomic sphere radius
to get sensibly converged results. Here we present results from calculations
on the basis of atomic sphere approximation (ASA) with
the ratio of the atomic sphere radius $r$ being
set to be $r\mbox{[R]}:r\mbox{[Fe]}:r\mbox{[B]}=1.3:1.0:0.9$
to adjust the number of electrons within each atomic sphere as precisely as possible.

\subsection{Target observables}
\label{sec::methods::observables}

All of the relevant intrinsic properties for a ferromagnet to make a permanent magnet
are calculated from first principles, following either or both of the approaches described above.
That is, formation energy and mixing energy
that will be denoted by $E_{\rm f}$ and $E_{\rm mix}$, respectively,
and magnetization, uni-axial magnetic anisotropy energy, and Curie temperature,
as denoted by $M$, $K$, and $T_{\rm Curie}$, respectively.

\subsubsection{Formation energy and mixing energy}

With the optimized structure for the target material and the reference elemental systems,
the formation energy for (La$_{1-x}$Ce$_{x}$)$_2$Fe$_{14}$B
is calculated as follows.
\begin{eqnarray*}
E_{\rm f}(x=0) & = & U_{\rm tot}[\mbox{La$_2$Fe$_{14}$B}]-2U_{\rm tot}[\mbox{dhcp-La}]/4  \\
&& -14U_{\rm tot}[\mbox{bcc-Fe}]-U_{\rm tot}[\mbox{$\alpha$-B}]/36 \\
E_{\rm f}(x=1) & = & 
U_{\rm tot}[\mbox{Ce$_2$Fe$_{14}$B}]-2U_{\rm tot}[\mbox{fcc-Ce}] \\
&& -14U_{\rm tot}[\mbox{bcc-Fe}]-U_{\rm tot}[\mbox{$\alpha$-B}]/36 
\end{eqnarray*}
Here $U_{\rm tot}[M]$ is the calculated energy for a given material $M$.
The unit cell of R$_2$Fe$_{14}$B contains four formula units and thus has 8 rare earth sites.
Half of them makes R($4f$) sublattices and the other half makes R($4g$) sublattices
and the latter occupies slightly larger space than the former in the notation of Herbst {\it et al}~\cite{rmp_1991,saito_2017}.
Replacing a La atom in La$_2$Fe$_{14}$B by Ce atom is doable and replacements of La by Ce is proceeded one by one
up to Ce$_2$Fe$_{14}$B with the step of 1/8 on the $x$ axis for (La$_{1-x}$Ce$_x$)$_2$Fe$_{14}$B.
For each of those discretely replaced materials the formation energy can be calculated
as follows.
\begin{eqnarray*}
&& E_{\rm f}(x=n/8) \\
 & = & U_{\rm tot}[\mbox{La$_{1-n/8}$Ce$_{n/8}$Fe$_{14}$B}] \\
 & & -2\left[
\left(1-\frac{n}{8}\right)\frac{U_{\rm tot}[\mbox{dhcp-La}]}{4} + \frac{n}{8}U_{\rm tot}[\mbox{fcc-Ce}]
\right] \\
 & & -14U_{\rm tot}[\mbox{bcc-Fe}]-\frac{U_{\rm tot}[\mbox{$\alpha$-B}]}{36}
\end{eqnarray*}
Here $n=0,1,2,\ldots,8$.

Analogous calculations can be done with KKR-CPA in principle but there is a slow
convergence problem with respect to the cutoff parameter $l_{\rm max}$ in KKR-CPA~\cite{zeller_2013}.
Within the present calculations where we fix $l_{\rm max}=3$ for La and Ce while the rest of the electronic states
are addressed with $l_{\rm max}=2$, the best we can do is to look at the mixing energy defined as follows.
\begin{eqnarray*}
&& E_{\rm mix}[\mbox{(La$_{1-x}$Ce$_{x}$)$_2$Fe$_{14}$B}] \\
 & = & U_{\rm tot}[\mbox{La$_{1-x}$Ce$_{x}$Fe$_{14}$B}] \\
 && -(1-x)U_{\rm tot}[\mbox{La$_2$Fe$_{14}$B}] - x U_{\rm tot}[\mbox{Ce$_2$Fe$_{14}$B}].
\end{eqnarray*}
The mixing energy can be calculated both by OpenMX and AkaiKKR and matched with each other.

\subsubsection{Magnetization}

Ferromagnetic ground state is reached with the {\it ab initio} structure optimization
calculations and the associated magnetization is extracted.
With the optimized cell volume, magnetization in Tesla is extracted as well.
This is the intrinsic magnetization of the interest in the context of permanent magnets.

While the chemical composition dependence of the lattice constants is incorporated
in the calculations in KKR-CPA, the interpolation is done only for the chemical composition.
At each of the discretely interpolated points with $x=n/8$ where $n=1,2,\ldots, 7$, effects
of segregation of doped Ce is incorporated in the structure optimization via OpenMX
while continuous interpolation both for the dopant concentration $x$ and site segregation ratio
is explored with KKR-CPA. The behavior of magnetization on the linearly interpolated lattice
and on the optimized lattice on each discretely optimized lattice would be overall
similar while inspection of some details tells us how the dopant Ce can bring
about an optimal magnetism as is discussed in Sec.~\ref{sec::discussions::volume}.

\subsubsection{Magnetic anisotropy energy at the stoichiometric limits}

At the stoichiometric limits $x=0$ and $x=1$, fully relativistic calculations with a constraint on
the direction of magnetization are done utilizing OpenMX
in order to inspect the dependence of the calculated energy
as a function of the angle between the crystallographic $c$-axis and the magnetization
to extract the magnetic anisotropy energy. Off-stoichiometric compounds on the basis of discrete
replacements may show substituted-site-dependent anisotropy which is not quite at the focus
of the present study.

With KKR-CPA as implemented in AkaiKKR, the magnetic anisotropy energy can be investigated
in the same way on the basis of the scalar-relativistic calculation
and taking into account the diagonal part
of the spin-orbit interaction~\cite{ogura_2015}. For the present purpose
we use the outputs from OpenMX to look at the fully relativistic effects.

\subsubsection{Curie temperature}
\label{sec::methods::observables::Tc}

Curie temperature is estimated on the basis of the ferromagnetic state
following Liechtenstein et al.~\cite{sasha_1987}
to derive an effective Heisenberg model for a simulation of finite-temperature magnetism.
Overestimates can come in two-fold: a) the problem
in describing the delocalized electronic state in the magnetism of intermetallics
on the basis of localized degrees of freedom~\cite{mm_2018}. b) assuming that
the spin Hamiltonian can be valid, if we do a mean field approximation the transition
temperature can typically be overestimated by a few tens of \%.
{\it Ab initio} data for Curie temperature is collected with KKR-CPA as implemented in AkaiKKR
on the basis of mapped Heisenberg model following Liechtenstein {\it et al.}
and the mean field approximation. For OpenMX
the practical implementation is getting to be done
and will be reported separately~\cite{terasawa}.

\subsection{Data integration}
\label{sec::methods::integration}

\subsubsection{Matching the data within {\it ab initio} approaches}

Summarizing Secs.~\ref{sec::methods::codes}~and~\ref{sec::methods::observables},
we work with a particular combination of the target property
and the approach to address the intrinsic properties of (La,Ce)$_2$Fe$_{14}$B
as is summarized in Table~\ref{table::combined}.
\begin{table}
\begin{tabular}{cccccc} \hline
 & $E_{\rm f}$ & $E_{\rm mix}$ & $M$ & $K$ & $T_{\rm Curie}$  \\ \hline
OpenMX  & $\circ$ & $\circ$ & $\circ$ & $\circ$ & $\star$  \\ \hline
AkaiKKR & \#\#      & $\circ$ & $\circ$ & \#      & $\circ$  \\ \hline
\end{tabular}
\caption{\label{table::combined} Our combined methods with {\it ab initio} structure optimization
employing the open-source software package OpenMX~\cite{OpenMX}
and KKR-CPA as implemented in AkaiKKR~\cite{AkaiKKR}.
The data are taken at the particular combination of {\it ab initio} code
and the target observable marked with $\circ$.
\#\# and \# is not used. $\star$ is being developed~\cite{terasawa}.
See the text for the rationale to take the particular combination
of the observable and the code.}
\end{table}

\subsubsection{{\it ab inito} and experimental data}

For the Curie temperature, we need to
refer to both of our calculated results with KKR-CPA
and also to the experimental data found in the literature~\cite{rmp_1991}.
This is because an estimated Curie temperature
on the basis of an effective spin model gives a significant overestimate~\cite{mm_2018}
due to the problem described above in Sec.~\ref{sec::methods::observables::Tc}.
In the present problem to assess the utility of a material
in a quantitatively given temperature range, unfortunately {\it ab initio} results by themselves
would not be quite sufficient. We need an offset imposed by experimental Curie temperature taken from the literature~\cite{rmp_1991}
to define a working merit function referring to a high-temperature edge in practical applications
of permanent magnets.

\section{Results}
\label{sec::results}

First we show {\it ab initio} data for the prerequisite properties
for (La,Ce)$_2$Fe$_{14}$B to qualify for a permanent magnet compound
and then assess their relative merits referring to the intrinsic properties
of the champion magnet compound, Nd$_2$Fe$_{14}$B, and the externally imposed
condition of usage concerning the typical operation temperatures.

\subsection{Calculated intrinsic properties}
\label{sec::data}

\subsubsection{Formation energy and mixing energy}
\label{sec::data::energy}

\begin{figure}
\scalebox{0.8}{\includegraphics{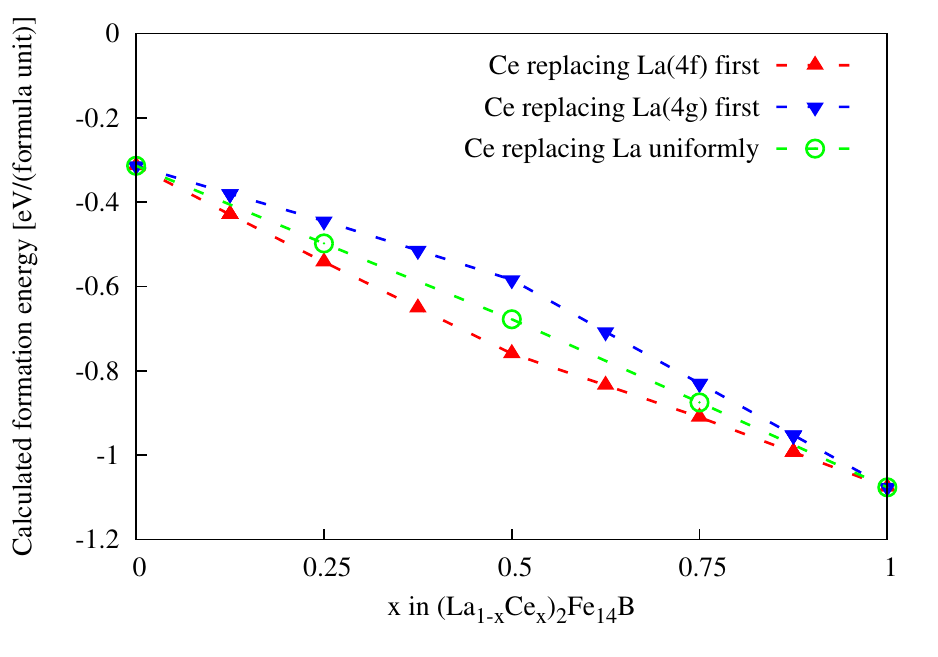}}
\caption{\label{fig::formation_energy} (Color online)
Calculated formation energy
of (La$_{1-x}$Ce$_{x}$)$_2$Fe$_{14}$B via {\it ab initio} structure optimization
utilizing OpenMX~\cite{OpenMX}.}
\end{figure}
Calculated formation energy for discretely substituted (La,Ce)$_2$Fe$_{14}$B is shown
in Fig.~\ref{fig::formation_energy} from {\it ab initio} structure optimization runs
utilizing OpenMX~\cite{OpenMX}. The formation energy is based on the calculation
of energy within the given standard pseudopotential data sets~\cite{OpenMX}.
Calculated energy results for the reference elemental systems are given in Table~\ref{table::elemental_references}
and those for the target stoichiometric compounds are in Table~\ref{table::target_materials}.
Other discretely substituted materials, with Ce either replacing La($4f$) first or La($4g$) first,
and substituting La($4f$) and La($4g$) on an equal footing (here
we need two Ce atoms for one step of substitution and thus the step on the $x$ axis is
$2~\mbox{(Ce atoms)}/8~\mbox{(host La sites)}=0.25$. The overall trend of Ce stabilizing
the crystal structure is clearly seen and La($4f$) is energetically preferred by Ce.
\begin{table}
\begin{tabular}{ccc} \hline
material & $N_{\rm atom}$ & $U_{\rm tot}~\mbox{[Hartree]}$  \\ \hline
$\alpha$-B & 36 & $-105.056$  \\ \hline
bcc-Fe & 1 & $-89.5730$  \\ \hline
dhcp-La & 4 & $-161.9402$  \\ \hline
fcc-Ce & 1 & $-53.9134$ \\ \hline
\end{tabular}
\caption{\label{table::elemental_references} Calculated energy
with OpenMX on the basis of the standard pseudopotentials~\cite{OpenMX}
and the choice of basis sets as described in Sec.~\ref{sec::methods}
for each elemental reference material.
$N_{\rm atom}$ is the number of atoms in the unit cell. For all of the data in this table,
the number of k points is $512$. Calculated energy for the reference elemental systems
$\alpha$-B and bcc-Fe
are taken from Ref.~\onlinecite{mm_20181228}.}
\end{table}
\begin{table}
\begin{tabular}{lcc} \hline
material  & $U_{\rm tot}~\mbox{[Hartree/(cell)]}$ & $E_{\rm f}~\mbox{[eV/(f.u.)]}$ \\ \hline
La$_2$Fe$_{14}$B & $-5351.689$ & $-0.32$ \\ \hline
Ce$_2$Fe$_{14}$B & $-5459.227$ & $-1.08$ \\ \hline\hline
\end{tabular}
\caption{\label{table::target_materials} Calculated energy with OpenMX
on the basis of the standard pseudopotentials and the choice
of basis sets as described in Sec.~\ref{sec::methods::OpenMX} and
the corresponding formation energy for each target material.}
\end{table}

\begin{figure}
\scalebox{0.8}{\includegraphics{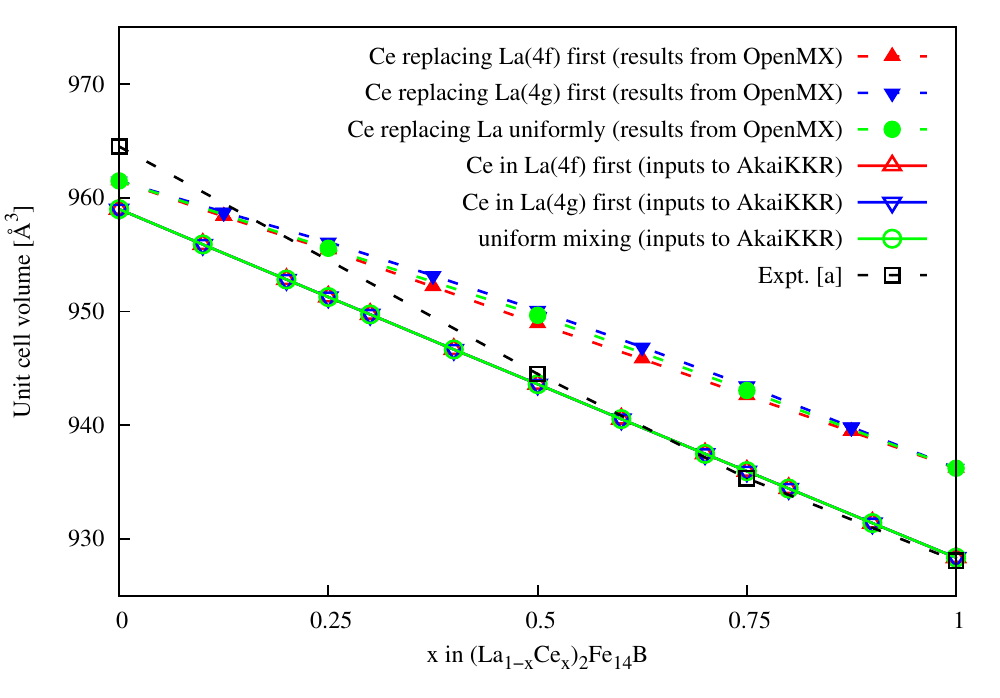}}
\caption{\label{fig::volume_comparison} (Color online)
Calculated unit-cell volume of (La$_{1-x}$Ce$_{x}$)$_2$Fe$_{14}$B via {\it ab initio} structure optimization,
the unit cell volume with the inputs to KKR-CPA on the
interpolated lattices referring to experimental
lattice in the literature~\cite{rmp_1991}, and results from
the latest experimental measurements~\cite{apl_2016}
as denoted by [a] in the figure.}
\end{figure}

Before moving on to KKR-CPA results for the mixing energy,
we compare the lattice as yielded from the structure optimization
and the inputs to KKR-CPA as we empirically defined
as is described in Sec.~\ref{sec::methods::AkaiKKR}.
As the representative parameter to characterize the lattice,
the data for the unit cell volume is shown in Fig.~\ref{fig::volume_comparison}
together with the recent experimental results~\cite{apl_2016}. We note
that the materials with Ce preferentially substituting La($4f$) sites
comes with a slightly smaller volume than the other cases
of uniform substitution or Ce preferentially substituting La($4g$) sites.
Such difference of unit cell volume depending on the site segregation
has not been taken into account in the KKR-CPA calculations
taking the interpolated lattice information based on past experimental data~\cite{rmp_1991}
as the input. Overestimate
of the unit cell volume is seen in the optimized lattice while the overall trend
with respect to Ce concentration
goes in parallel between all of the optimized lattice, interpolated lattice
in the input to KKR-CPA, and recent experimental measurement~\cite{apl_2016}.
Larger deviation seen at La$_2$Fe$_{14}$B between the recent experimental data,
optimized lattice, and the past experiments~\cite{rmp_1991}
might be related to the less robust structure of La$_2$Fe$_{14}$B as seen
in the relatively small absolute value
of the calculated formation energy in Table~\ref{table::target_materials}
and Fig.~\ref{fig::formation_energy}.

\begin{figure}
\begin{tabular}{l}
(a) \\
\scalebox{0.8}{\includegraphics{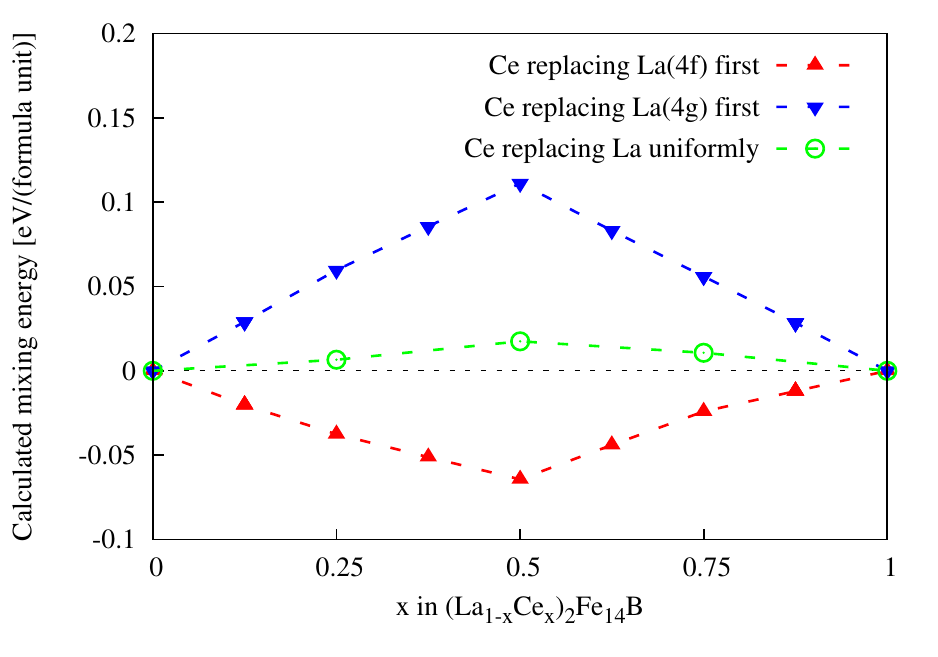}}\\
(b) \\
\scalebox{0.8}{\includegraphics{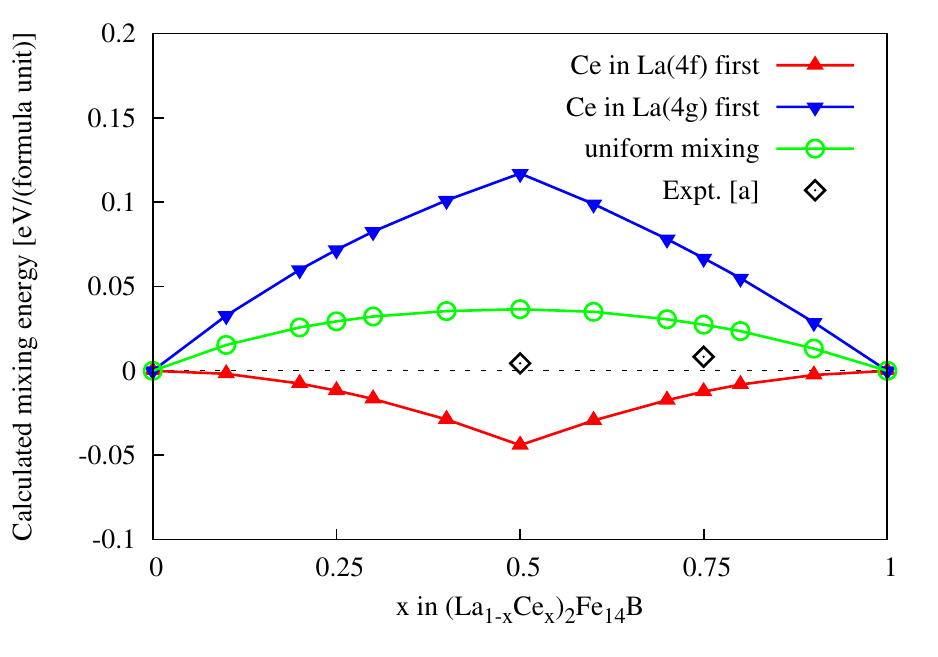}}
\end{tabular}
\caption{\label{fig::mixing_energy} (Color online)
Calculated mixing energy 
for (La$_{1-x}$Ce$_{x}$)$_2$Fe$_{14}$B. (a) Results from the {\it ab inito} structure optimization utilizing OpenMX and
(b) KKR-CPA utilizing AkaiKKR where results corresponding
to the latest experimentally measured site segregation~\cite{apl_2016}
are denoted by [a] in the figure.}
\end{figure}

Now we match the formation energy from the structure optimization
and the other data from KKR-CPA. Those two data sets are presented
in Fig.~\ref{fig::mixing_energy}. We see that the results from the
two approaches agree semi-quantitatively. Even with the small difference
in the optimized cell volume depending on the site segregation of dopant Ce,
we can presume that OpenMX data and AkaiKKR data are going in parallel
concerning the energetics.

\subsubsection{Magnetization}
\label{sec::data::magnetization}

\begin{figure}
\begin{tabular}{l}
(a) \\
\scalebox{0.8}{\includegraphics{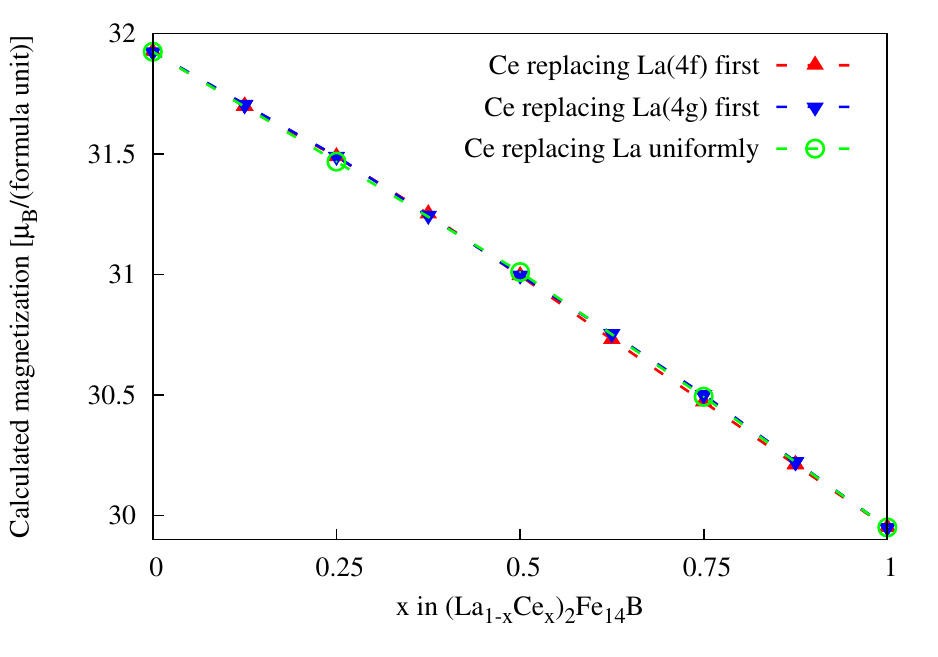}} \\
(b) \\
\scalebox{0.8}{\includegraphics{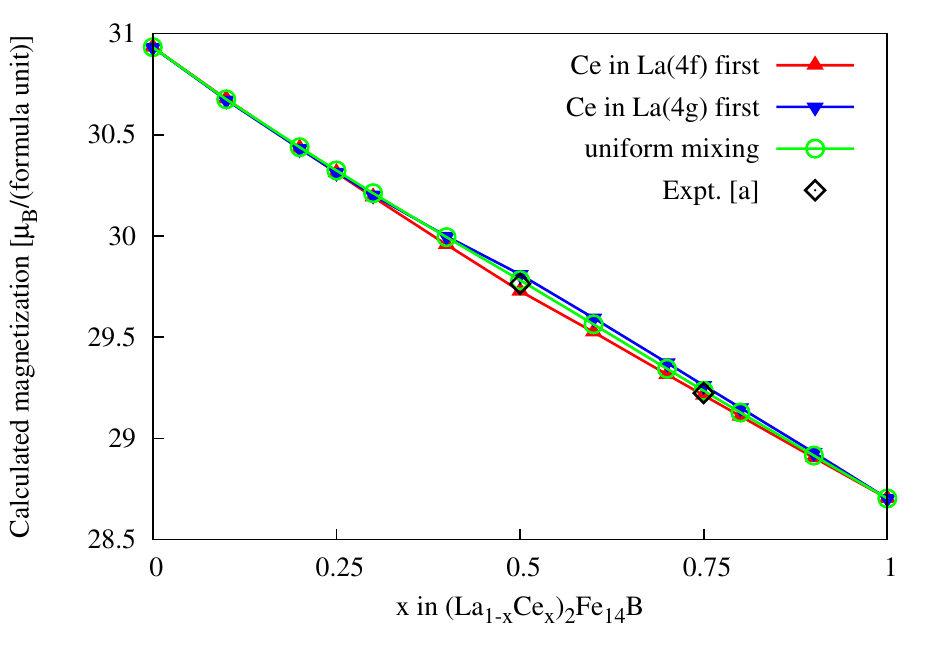}} 
\end{tabular}
\caption{\label{fig::mag} Calculated magnetization from (a) {\it ab initio} structure optimization and (b) KKR-CPA
on the interpolated lattice between La$_2$Fe$_{14}$B and Ce$_{2}$Fe$_{14}$B.
Results corresponding to the latest experimentally measured site segregation~\cite{apl_2016}
are denoted by [a] in the figure.
}
\end{figure}

\begin{figure}
\begin{tabular}{l}
(a) \\
\scalebox{0.8}{\includegraphics{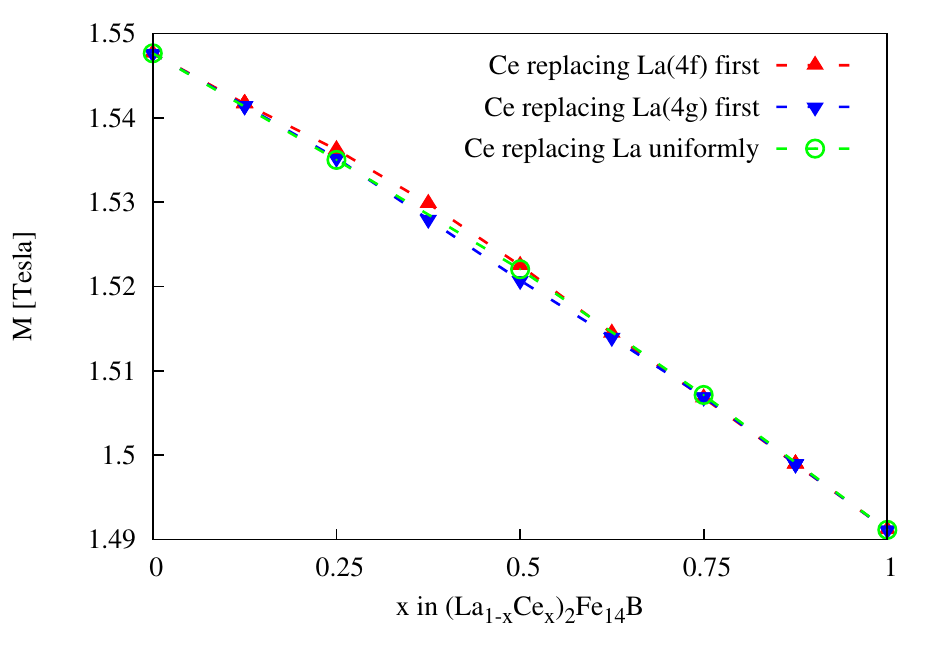}}\\
(b) \\
\scalebox{0.8}{\includegraphics{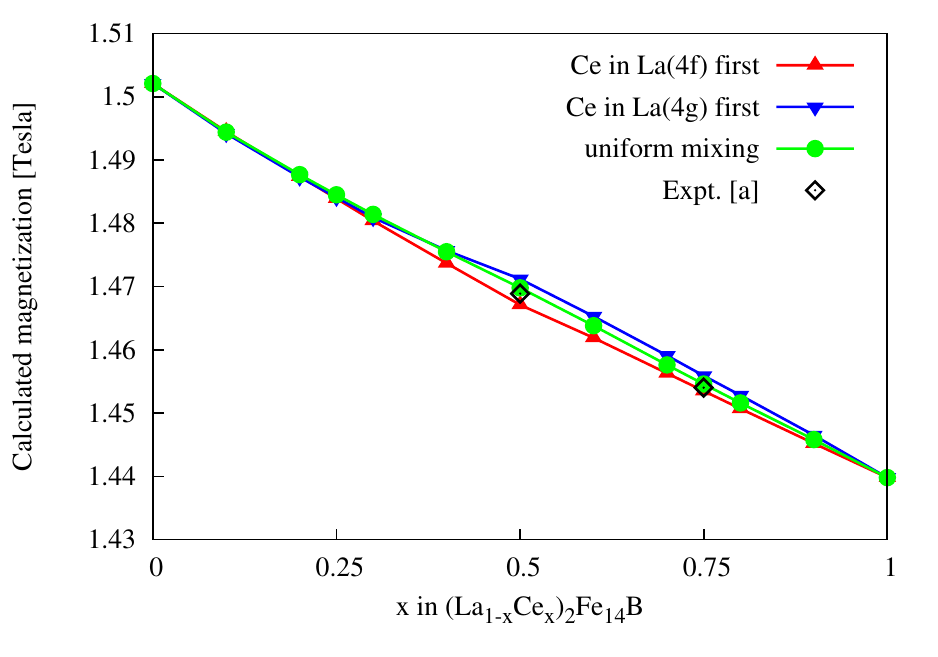}}
\end{tabular}
\caption{\label{fig::mag_in_Tesla} Calculated magnetization in Tesla
from (a) {\it ab initio} structure optimization and (b) KKR-CPA
on the interpolated lattice between La$_2$Fe$_{14}$B and Ce$_{2}$Fe$_{14}$B.
Results corresponding to the latest experimentally measured site segregation~\cite{apl_2016}
are denoted by [a] in the figure.
}
\end{figure}

Calculated magnetization of (La$_{1-x}$Ce$_{x}$)$_2$Fe$_{14}$B is shown in Fig.~\ref{fig::mag}
as magnetic moments per formula unit. Taking into account the volume as shown in Fig.~\ref{fig::volume_comparison},
the trend of magnetization in Tesla looks like Fig.~\ref{fig::mag_in_Tesla}. The latter should be used
in the evaluation of the merit for permanent magnets.

In the results of structure optimization via OpenMX,
we see that the small volume effect with Ce preferentially
replacing La($4f$) sites as remarked in Sec.~\ref{sec::data::energy} is reflected in the slight supremacy of magnetization in Tesla
of the cases with Ce preferentially replacing La($4f$) sites
over other case with Ce preferentially replacing La($4g$) sites. Remarkably,
the energetically favorable substitution coincides with the case where a stronger magnetization in Tesla
is reached due to the volume effect. This is a rare trend and is in contrast to typical situations where
strong magnetization and structure stability is often traded off.

On the other hand KKR-CPA results via AkaiKKR do not incorporate the volume effect
that can come from the doped site segregation, but the mixing ratio can be continuously
swept on demand to see the overall trends including the experimental data points.
While we systematically explored three representative case studies with
uniform replacements and Ce preferentially replacing La($4f$) or La($4g$) sites,
fractional ratio in the replacements was inspected in the recent experiments~\cite{apl_2016}.
Simulating such experimental mixing ratio within KKR-CPA, calculated magnetization
is plotted in Fig.~\ref{fig::mag}~(b)~and~\ref{fig::mag_in_Tesla}~(b). It is seen
that while energetically favorable replacements of La($4f$) by Ce dominates
slightly shifted distribution over to the other energetically unfavorable La($4g$) site
helps in lifting up the magnetization within the fixed lattice.

Since we do not exactly simulate the experimental observation at the moment
it is not clear which physics of the above would be more dominating in real (La,Ce)$_2$Fe$_{14}$B.
At least it is clear the particular site segregation is important in getting magnetization
beyond a plainly interpolated magnetization between La$_2$Fe$_{14}$B and Ce$_2$Fe$_{14}$B.

\subsubsection{Magnetic anisotropy energy}
\label{sec::data::MAE}

\begin{figure}
\scalebox{0.8}{\includegraphics{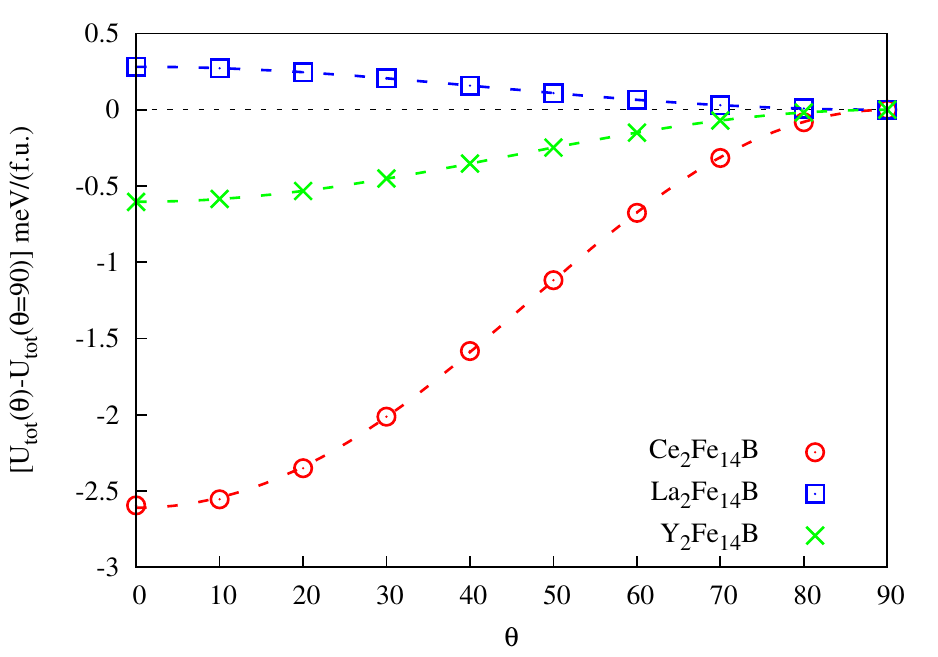}}\\
\caption{\label{fig::MAE} (Color online) Calculating magnetic anisotropy energy via fully relativistic calculation
as implemented in OpenMX. Lines are fits to the calculated data points.}
\end{figure}
Results from fully relativistic calculations utilizing OpenMX
with constraints on the direction of magnetization for stoichiometric compounds are shown in
Fig~\ref{fig::MAE}. As a reference Y$_2$Fe$_{14}$B is included for which the structure optimization
to extract the formation energy is presented in Ref.~\onlinecite{mm_20181228}.
Optimized structure is taken from Sec.~\ref{sec::data::energy} for La$_2$Fe$_{14}$B and Ce$_2$Fe$_{14}$B
and the constraint on the direction of magnetization,
as measured by an angle from the crystallographic $c$-axis of the R$_2$Fe$_{14}$B crystal~\cite{rmp_1991},
is imposed to estimate the energy within the pseudopotential and the choice of the basis sets described in Sec.~\ref{sec::methods::OpenMX}.

In Fig.~\ref{fig::MAE} calculated energy $\left.U_{\rm tot}[\mbox{R$_2$Fe$_{14}$B}]\right|_{\theta}$ is plotted
with an offset taken at $\theta=90~\mbox{degrees}$. The fit to the calculated data obtained at every 10 degrees
from $\theta=0$ to $\theta=90~\mbox{degrees}$ with the following relation
\begin{equation}
-K\left[(1-p-q)\cos^2\theta+p\cos^4\theta+q\cos^6\theta\right]
\label{eq::aniso}
\end{equation}
gives the anisotropy energy and the higher order contribution
as tabulated in Table~\ref{table::aniso}.
\begin{table}
\begin{tabular}{cccc} \hline
material & $K~\mbox{[meV/(f.u.)]}$ & $p$ & $q$ \\ \hline
Ce$_2$Fe$_{14}$B & $2.609(7)$ & $0.14(6)$ & $-0.15(4)$ \\ \hline 
La$_2$Fe$_{14}$B & $-0.2819(1)$ & $0.097(1)$ & $q \equiv 0$ \\ \hline
Y$_2$Fe$_{14}$B & $0.6034(2)$ & $0.009(1)$ & $q\equiv 0$ \\ \hline
\end{tabular}
\caption{\label{table::aniso} Calculated
magnetic anisotropy constants as defined in Eq.~(\ref{eq::aniso}) in the text.
The number in parenthesis is the error in the last digit from the data fit.}
\end{table}

Numerically observed trends for uni-axial magnetic anisotropy
are qualitatively
consistent with the experimentally reported trends found in the literature~\cite{expt_1985}
where the strength of uni-axial magnetic anisotropy follows the order
\[
\mbox{Ce$_2$Fe$_{14}$B} > \mbox{Y$_2$Fe$_{14}$B}  > \mbox{La$_2$Fe$_{14}$B}.
\]
We find easy-plane anisotropy in La$_{2}$Fe$_{14}$B which is by itself
not entirely consistent with the experiments~\cite{expt_1985}.
Presumably in our calculations $3d$-electron anisotropy is underestimated
and calculated anisotropy for La$_2$Fe$_{14}$B has not been
strong enough but it is to be noted that actually
La seems to contribute to the easy-plane anisotropy in La$_2$Fe$_{14}$B.
It is also remarkable that higher-order terms up to the $3^{rd}$ order
from delocalized $4f$-electrons in Ce$_2$Fe$_{14}$B and the $2^{nd}$ order terms
in $d$-electron anisotropy can be quantitatively determined from first principles.

Uni-axial magnetic anisotropy energy contributed from itinerant $4f$-electrons
in Ce$_{2}$Fe$_{14}$B are estimated to be in the order of 1~meV from 2 Ce atoms
as seen in Fig.~\ref{fig::MAE}. This is not as large as the typical scale
of MAE from RE like Nd, Sm, and Dy amounting to the order of 10~meV while an order
of magnitude larger than the $3d$-electron anisotropy in the order of 0.1~meV per atom
at maximum. Thus contribution to anisotropy coming from Ce in Ce-Fe intermetallics
should definitely be exploited in fabricating an intermediate-grade magnet.

\subsubsection{Curie temperature}

\begin{figure}
\scalebox{0.8}{\includegraphics{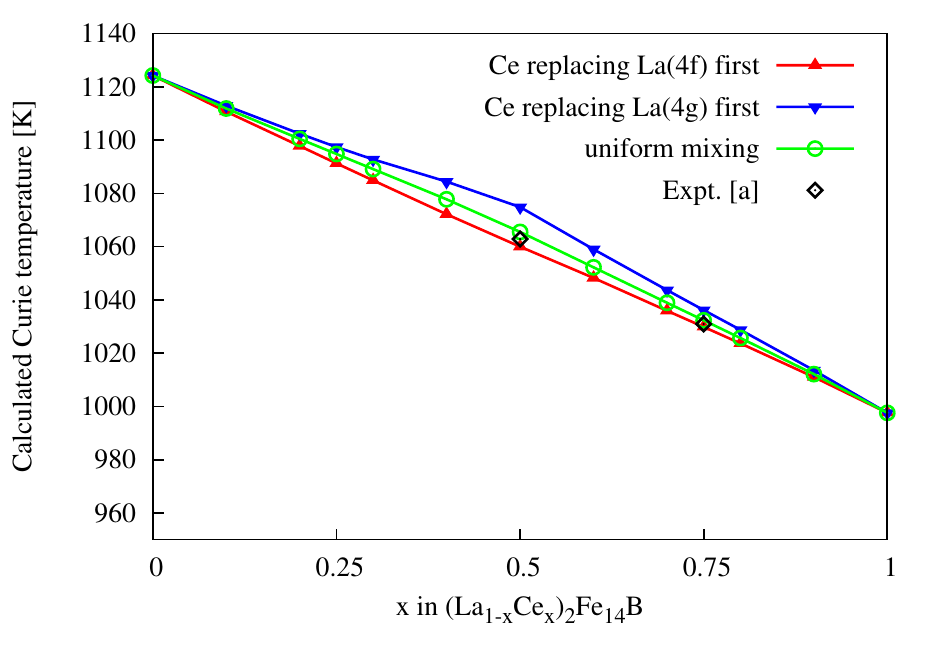}}\\
\caption{\label{fig::Tc} (Color online) Calculated Curie temperature for the interpolated-lattice compounds
as addressed via KKR-CPA.
Results corresponding to the latest experimentally measured site segregation~\cite{apl_2016}
are denoted by [a] in the figure.}
\end{figure}

Calculated Curie temperatures are shown in Fig~\ref{fig::Tc}
from KKR-CPA calculations.
At the stoichiometric limits, experimental Curie temperatures are~\cite{rmp_1991}
$T_{\rm Curie}[\mbox{La$_2$Fe$_{14}$B}]=530~\mbox{[K]}$
and 424~K for Ce$_{2}$Fe$_{14}$B. An overestimate for the absolute value of the Curie temperature
by a factor close to $2$ is seen presumably due to the origins described in Sec.~\ref{sec::methods::observables::Tc}
while the slope between La$_2$Fe$_{14}$B and Ce$_{2}$Fe$_{14}$B
spanning $100~\mbox{[K]}$ is approximately reproduced.

\subsection{Assessment of the merit of the light-rare-earth magnet}

\subsubsection{Referring to Nd$_2$Fe$_{14}$B}

The intrinsic properties calculated above are compared to the counterpart data
for Nd$_2$Fe$_{14}$B and the relative merit of (La$_{1-x}$Ce$_x$)$_2$Fe$_{14}$B is assessed
to inspect the optimal $x$.
\begin{figure}
\scalebox{0.8}{\includegraphics{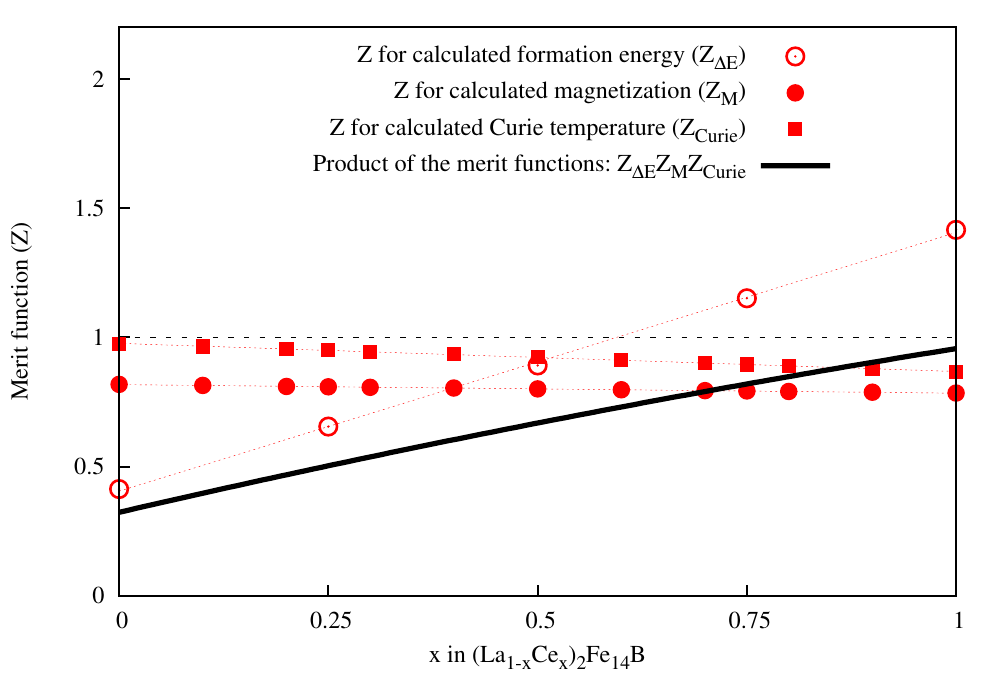}}
\caption{\label{fig::calc_merit}
(Color online) Relative merit of (La$_{1-x}$Ce$_x$)$_2$Fe$_{14}$B measured with respect to Nd$_2$Fe$_{14}$B
for the formation energy, magnetization in Tesla, and Curie temperature based on our {\it ab initio} data.}
\end{figure}
We simply normalize the calculated formation energy, magnetization in Tesla, and Curie temperature
of (La$_{1-x}$Ce$_x$)$_2$Fe$_{14}$B
by the counterpart data for Nd$_2$Fe$_{14}$B. Because the overall trend of the intrinsic properties
as a function of $x$ more or less behaves overall in the same way irrespectively of the details of site
segregation, we focus on the calculations with uniform replacements between La and Ce. 
For the formation energy, $\Delta E_{\rm f}[\mbox{Nd$_2$Fe$_{14}$B}]
=-0.76~\mbox{[eV/(formula unit)]}$ is taken from Ref.~\onlinecite{mm_20181228} and define this ratio
as the merit of formation energy which measures the merit in structure stability in terms of energetics.
Also calculated magnetization in Tesla is compared to the counterpart results for Nd$_2$Fe$_{14}$B~\cite{mm_2018,mm_20181228}.
For Nd$_2$Fe$_{14}$B in the ground state $M_{s}=1.836~\mbox{[Tesla]}$ to which our {\it ab initio} calculated magnetization in the ground state is referred.
Concerning the Curie temperature of Nd$_2$Fe$_{14}$B,
the same set-up of KKR-CPA as is done for (La,Ce)$_2$Fe$_{14}$B
gives $T_{\rm Curie}=1152.743~\mbox{[K]}$, which should be carefully compared
with the experimental data $T_{\rm Curie}=585~\mbox{[K]}$~\cite{rmp_1991}.
Thus defined merit of (La$_{1-x}$Ce$_x$)$_2$Fe$_{14}$B concerning formation energy, magnetization,
and Curie temperature is shown in Fig.~\ref{fig::calc_merit}.
It is seen that the merit is gained mostly due to the formation energy and monotonically
increasing with respect to $x$ based on the plain definition of the merit function
only referring to Nd$_2$Fe$_{14}$B within {\it ab initio} data.

Since all of the intrinsic properties are prerequisites,
we have taken a product of the merit function for each
of formation energy, magnetization, and Curie temperature
rather than taking a sum.
It is not very straightforward to define the merit
of magnetic anisotropy energy on our {\it ab initio} data
because we have the data only for the stoichiometric limits.
On top of that,
a data set for anisotropy energy calculated
for Nd$_2$Fe$_{14}$B would be
based on the effects
of crystal fields acting on very well localized $4f$ electrons.
The mechanism of magnetic anisotropy is different between (La,Ce)$_2$Fe$_{14}$B and Nd$_2$Fe$_{14}$B.
It is not quite straightforward to take a data set for magnetic anisotropy on an equal footing
for the materials with delocalized $4f$-electrons and for the other bunch of materials with
very well localized $4f$-electrons. We will not take a very close look at anisotropy here
due to this fundamental problem and practically due to the trend
seen in Fig.~\ref{fig::calc_merit} which is monotonically-increasing with respect to $x$.
Here further inclusion of magnetic anisotropy energy would not qualitatively alter the message of the merit.
Rather the range of the practical operation temperature seems to put the most stringent
constraint on the merit function and make it non-monotonic
as we see below in Sec.~\ref{sec::the_operation_temperature_issue}.

\subsubsection{Considering the operation temperature range}
\label{sec::the_operation_temperature_issue}

\begin{figure}
\scalebox{0.8}{\includegraphics{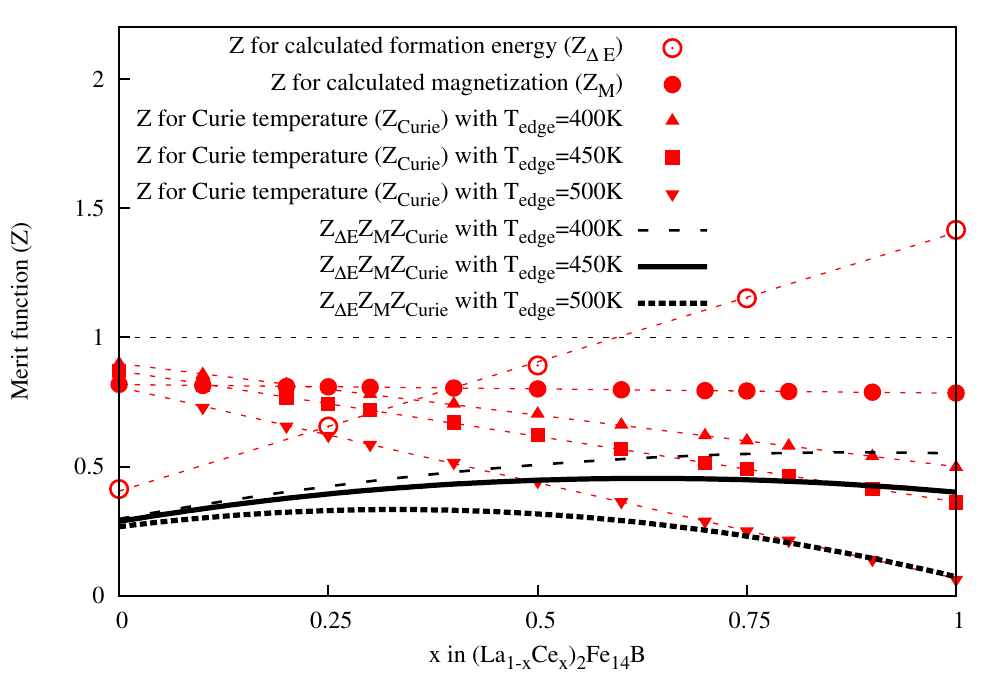}}
\caption{\label{fig::calc_merit_var}
(Color online) Relative merit of (La$_{1-x}$Ce$_x$)$_2$Fe$_{14}$B measured with respect to
Nd$_2$Fe$_{14}$B. The overall merit involving formation energy, magnetization, and Curie temperature
is quantitatively controlled by the position of the upper temperature limit $T_{\rm edge}$
of practical usage. Merit for three cases, $T_{\rm edge}=400~\mbox{K}$, $450~\mbox{[K]}$,
and $500~\mbox{[K]}$ are presented.}
\end{figure}
Permanent magnets are basically put into practical use
around room temperatures and the critical interest lies in the magnetic properties
at the high-temperature edge of the temperature range: $300~\mbox{[K]}\le T\le T_{\rm edge}~\mbox{[K]}$.
For traction motors in vehicles $T_{\rm edge}\simeq 450~\mbox{[K]}$.
A ferromagnet with the Curie temperature below $T_{\rm edge}$ would not be acceptable in practice.
Thus we define the merit function concerning the Curie temperature,
which we will denote by $Z_{\rm curie}$ below,
in such a way that the merit becomes unity at the Curie temperature of Nd$_2$Fe$_{14}$B
and zero at $T_{\rm Curie}(x)=T_{\rm edge}$. A way to define this would be the following:
\begin{eqnarray}
Z_{\rm Curie}(x) & \equiv & \frac{T^{\rm adjusted}_{\rm Curie}(x)}{T^{\rm expt.}_{\rm Curie}[\mbox{Nd$_2$Fe$_{14}$B}]} \nonumber \\
 & & \times \frac{T_{\rm Curie}^{\rm adjusted}(x)-T_{\rm edge}}{T_{\rm Curie}[\mbox{Nd$_2$Fe$_{14}$B}]-T_{\rm edge}}\\
T^{\rm adjusted}_{\rm Curie}(x) & \simeq & T^{\rm calc}_{\rm Curie}(x)
\frac{T^{\rm expt.}_{\rm Curie}[\mbox{Nd$_2$Fe$_{14}$B}]}{T^{\rm calc.}_{\rm Curie}[\mbox{Nd$_2$Fe$_{14}$B}]}
\end{eqnarray}
It is plotted in Fig.~\ref{fig::calc_merit_var}
incorporating the experimental data $T^{\rm expt.}_{\rm Curie}[{\rm Nd_2Fe_{14}B}]=585~\mbox{[K]}$~\cite{rmp_1991}
and picking up several choices for $T_{\rm edge}$ between $400~\mbox{[K]}$ and $500~\mbox{[K]}$.
The higher (lower) the operation temperature range spans, less (more) amount of Ce gives a more reasonable choice,
respectively. We see that when the high temperature edge is set at 450~K,
the merit function shows a broad maximum around 70\% of Ce.
This particular optimal chemical composition may be compared with the messages
from the recent developments for Ce-based core-shell magnet~\cite{ito_2016} which points to an optimal amount
of Ce to be 75\% as investigated with the room temperature performance~\cite{shoji_2018}.

The above arguments depend on how we define the merit function up to the location of the high temperature edge.
The merit function can be further generalized via possible data assimilation
to incorporate extrinsic parameters, even the cost of the fabrication.
In the present study we have restricted ourselves to an evaluation of the intrinsic properties of magnetism.

\section{Discussions}
\label{sec::discussions}

\subsection{Utility of Ce}
\label{sec::discussions::volume}

Obvious merit brought about by Ce has been in the structure stabilization
and the biggest drawback was coming from Curie temperature.
During the course of our calculations we notice the tiny enhancement of magnetization
as measured in Tesla in the results of {\it ab initio} structure optimization
as shown in Fig.~\ref{fig::mag_in_Tesla}~(a). Even though the magnetization
as measured in terms of magnetic moments per atoms may be on a par as seen in Fig.~\ref{fig::mag}~(a),
the particular volume shrinkage can help in enhancing the merit of the material.
Even if $4f$-electrons may not contribute to the intrinsic magnetism explicitly
these indirect help as a useful spacer should not be overlooked.

\subsection{Valence state of Ce}
\label{sec::discussions::valence}

In the present work we have assumed that Ce stays in the tetravalent state.
On the other hand, trends toward trivalent state of Ce in an expanded space may well be suspected,
especially when Ce replaces the slightly larger La($4g$) site. Generally speaking,
it seems unlikely for the $4f$ electron state of Ce to be able to remain localized
if the conduction band is widely exchange split. The location of localized $4f$-level in Ce
is around 2~eV below the Fermi level while the exchange splitting can span a range
of few electron volts. Thus it is naively expected that in Ce-Fe intermetallic ferromagnets
$f$-$d$ hybridization would be too strong to allow for localized $4f$-electrons in Ce
unless some fine tuning of particular electronic clouds is implemented for some mechanism
possibly exploiting magnetic anisotropy. At the moment we are not aware of such cases
but do not entirely rule out the possibility for trivalent state in Ce-Fe (or Ce-Co) intermetallics.
Separate calculations assisted by a dynamical mean field theory to
describe the correlated electron nature in Ce
for a fictitious Ce$_{2}$Cu$_{14}$B on the La$_2$Fe$_{14}$B lattice~\cite{mm},
which is constructed so that the localization of $4f$ electron in Ce would get more likely
with the largest lattice constant and vanishing splitting of the conduction electrons,
have showed no sign of trivalent Ce. Presumably the lattice structure of R$_2$Fe$_{14}$B
may not be optimal for the localization of $4f$-electrons in Ce.

\section{Conclusions and outlook}
\label{sec::conclusions}

We have figured out the merit of light-rare-earth permanent magnet
(La,Ce)$_2$Fe$_{14}$B from first principles by incorporating the prerequisite
condition for practical utility referring to the experimental Curie temperature.
With the commonplace high temperature edge at 450~K in practical applications of rare earth permanent magnets,
the best compromise has been found
at the concentration of Ce around 70\%.

We have purged Co out of the present scope to clarify
the $4f$-electron physics in the valence state
combined with $3d$-electron ferromagnetism - this has been
to assess the maximum possible merit of ferromagneism
from first principles with abundant elements. Inclusion of Co does help at least
in raising the Curie temperature~\cite{rmp_1991}
as long as the extra cost for Co would not be too big a problem. Exploration of the extended
chemical composition space for R$_{2}$(Fe,Co)$_{14}$B (R=rare earth including La and Ce) in quest
of peak performance in uni-axial ferromagnetism
will be reported in a separate work~\cite{harashima}.

\begin{acknowledgments}
MM's work in ISSP, University of Tokyo
is supported by Toyota Motor Corporation. Helpful comments given by T.~Ozaki and F.~Ishii
for {\it ab initio} calculations with OpenMX and
discussions with Y.~Harashima, T.~Miyake, K.~Tamai, N.~Kawashima,
M.~Hoffmann, A.~Ernst, A.~Marmodoro, S.~Mankovsky, and H.~Ebert in related projects
are gratefully acknowledged. Part of the present project was supported by JSPS KAKENHI Grant~No.~15K13525.
Numerical calculations were done on ISSP supercomputer center, Univ. of Tokyo.
\end{acknowledgments}

\end{document}